\documentclass[aps,prl,twocolumn]{revtex4-2}
\usepackage{amssymb}
\usepackage{amsmath}
\usepackage{graphicx}
\usepackage{color}

\begin{document}
\title{Resonance fluorescence and indistinguishable photons \\from a coherently driven B centre in hBN}
\author{Domitille G\'erard$^{1}$, St\'ephanie Buil$^{1}$, Kenji Watanabe$^{2}$, Takashi Taniguchi$^{3}$, Jean-Pierre Hermier$^{1}$, Aymeric Delteil$^{1}$}

\affiliation{ $^1$ Universit\'e Paris-Saclay, UVSQ, CNRS,  GEMaC, 78000, Versailles, France. \\
$^2$ Research Center for Electronic and Optical Materials, National Institute for Materials Science, 1-1 Namiki, Tsukuba 305-0044, Japan \\
$^3$ Research Center for Materials Nanoarchitectonics, National Institute for Materials Science,  1-1 Namiki, Tsukuba 305-0044, Japan \\ {\color{white}--------------------} aymeric.delteil@usvq.fr{\color{white}--------------------} }


\begin{abstract}
Optically active defects in hexagonal boron nitride (hBN) have become amongst the most attractive single-photon emitters in the solid state, owing to their high-quality photophysical properties, combined with the unlimited possibilities of integration offered by the host {van der Waals} material. In particular, the B centres, with their narrow linewidth, low wavelength spread and controllable positioning, have raised a particular interest for integrated quantum photonics. However, to date, either their excitation or their detection has been performed non-resonantly due to the difficulty of rejecting the backreflected laser light at the same wavelength, thereby preventing to take full benefit from their high coherence in quantum protocols. Here, we make use of narrow-linewidth emitters integrated in a hybrid metal-dielectric structure to implement cross-polarisation laser rejection. This allows us to observe resonantly scattered photons, with associated experimental signatures of optical coherence in both continuous-wave (cw) and pulsed regimes, respectively the Mollow triplet and Hong-Ou-Mandel interference from zero-phonon-line emission. The {two-photon interference visibilities  of $0.93 \pm 0.21$ and $0.92 \pm 0.26$ we measured for two emitters demonstrate} the potential of B centres in hBN for applications to integrated quantum information.
\end{abstract}


 \maketitle
Optically active artificial atoms in the solid state, such as self-assembled quantum dots (QDs) and crystal defects, are widely considered as key players in the emerging photonic quantum technologies, combining high-quality single-photon emission properties and possibilities of integration~\cite{Aharonovich16}. In such systems, optimal single-photon properties -- such as purity and indistinguishability -- can be obtained by performing both laser excitation and photon detection resonantly with the electronic transition of interest. Resonant excitation enables coherent control of the electronic states, and minimises excitation jitter and detrimental effects of the environment on the electronic states. In turn, detection of resonant photons is necessary for experiments based on photon interference. Indeed, only the fraction of photons that are free from phonon-assisted processes exhibit a high degree of indistinguishability. Implementing such resonance fluorescence has enabled pivotal demonstrations in photonic quantum information~\cite{He13} and quantum networks~\cite{Bernien13,Delteil15}.

\begin{figure*}[t]
  \centering
  \includegraphics[width=0.9\linewidth]{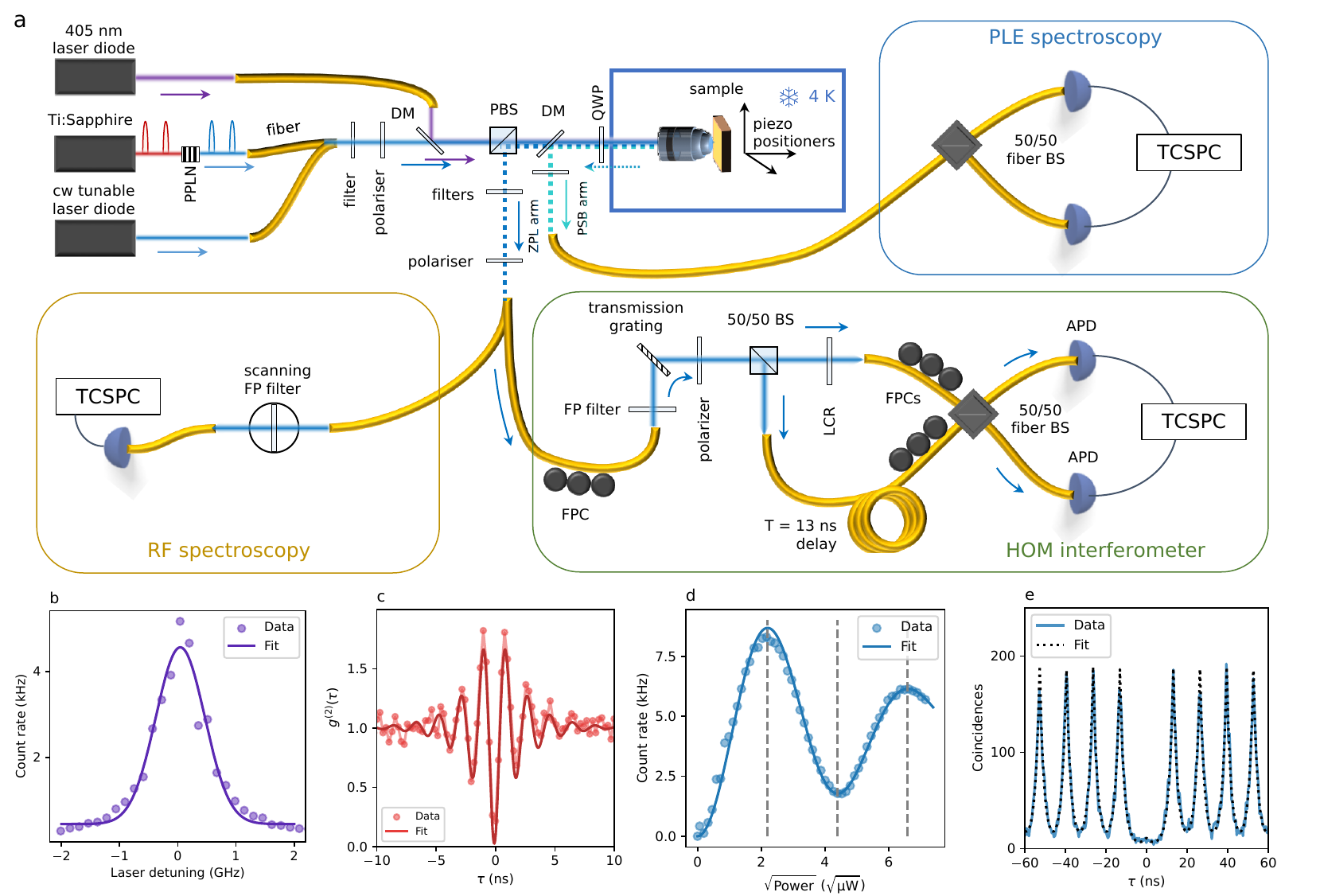}\\
  \caption{\textbf{Experimental setup and PLE characterisation of the emitter.} (a) Experimental setup. (b) Dots: low-power (40~nW) laser scan to infer the emitter inhomogeneous linewidth. Plain line: Gaussian fit to the data. (c) Dots: medium-power (0.5~$\mu$W) $g^{(2)}(\tau)$ revealing Rabi oscillations. Plain line: fit to the data. (d) Dots: count rate as a function of the square root of the laser power in pulsed regime. Plain line: damped sine fit to the data. (e) Blue curve: $g^{(2)}(\tau)$ in pulsed regime. Dotted line: {multi-peak} fit to the data.}  \label{fig1}
\end{figure*}

Most realisations to date are based on single-photon emitters (SPEs) hosted in 3D matrices, such as III-V QDs and NV centres. However, driven by the numerous possibilities of integration of the emerging {van der Waals} materials, free from lattice matching constraints and controlled down to the monolayer scale, a growing interest has been directed towards more recently discovered single-photon sources in {van der Waals} crystals such as transition metal dichalcogenides~\cite{chakraborty15, he15, koperski15, srivastava15, tonndorf15} and hexagonal boron nitride (hBN)~\cite{tran16}. Amongst the wide variety of optically active defects in the latter material, the so-called B centres stand out owing to their ability to be locally and reproducibly created in high quality, carbon-rich hBN crystals~\cite{Fournier21, Gale22, Shevitski19, Horder22} {down to 3~nm thickness~\cite{Liu25}}. { Their microscopic structure is currently understood as an out-of-plane carbon dimer in the split interstitial configuration, fabricated by electron-beam-induced conversion of ubiquitous UV-emitting in-plane dimers~\cite{Hou25, Nunez25, Plo25}. Their emission wavelength has a narrow (0.5~nm full width at half maximum) ensemble distribution around 436~nm~\cite{Fournier21}. Although this wavelength is not ideal for quantum communications, the controlled positioning of the emitters enables the development of integrated on-chip quantum computing devices.} Amongst the fascinating properties of these blue emitters, recent studies have unveiled their high coherence, with negligible dephasing under resonant excitation~\cite{Fournier23PRB,Gerard25}. However, the experiments yielding these conclusions have relied on the detection of phonon-assisted emission -- redshifted by up to 200~meV -- to allow for efficient laser rejection based on filtering{~\cite{Tran17}}. While such photoluminescence excitation (PLE) technique makes it possible to efficiently probe the electronic coherence, the latter does not profit to the collected photons, which are incoherent. The collection of resonant light scattering, or resonance fluorescence (RF), is notoriously difficult in confocal microscopy due to the simultaneous collection of backscattered laser light at the same energy, and requires excellent emitters in terms of brightness, linewidth, spectral diffusion and saturation power, together with complex rejection techniques, based on spatial mode mismatch or cross-polarisation (dark field) techniques~\cite{Kuhlmann13, Benelajla21}. In the solid state, such techniques have been successfully applied to self-assembled QDs~{\cite{Muller07, Vamivakas09}}, which benefit from a low saturation power, narrow linewidth and possibilities of integration in photonic structures.

In {van der Waals} materials, recent works have shown detection of scattered light close to the excitation laser wavelength~\cite{Konthasinghe19, Errando21}, yet without compelling evidence that the collected light is indeed RF from the zero-phonon-line (ZPL) and not close-to-resonance but incoherent light (involving \textit{e.g.} low-energy acoustic phonons in either the excitation or the emission process). The difficulty to observe RF in hBN has both physical and technical origins. On the one hand, the relevant optical properties of the SPEs are less favourable as compared to QDs -- in particular, they typically have a higher saturation power, thus requiring a larger laser power to reach a given photon count rate. Their dephasing rate and spectral diffusion are also often sizeably higher, further degrading the signal over laser background ratio. Additionally, the free-space optical components required to implement RF, such as polarisers and birefringent crystals, typically have lower performance in their emission range (in particular in the blue range) than their near infrared counterparts.

Here, we implement RF on single B centres generated by an electron beam in {170~nm and 220~nm} thick hBN crystals on top of a silver mirror (see Methods). The latter allows for an enhanced collection efficiency while maintaining a planar configuration, favourable for implementing polarisation suppression. The sample is inserted in a low-temperature confocal microscope based on a close-cycle cryostat operating at 3.5~K. The experimental setup is depicted on fig.~\ref{fig1}a. The emitter can be excited in both cw and pulsed regimes. A 405~nm laser diode is used to perform photoluminescence (PL), while resonant excitation is realised with either a cw laser diode or a mode-locked, frequency-doubled Ti:Sapphire laser yielding 15~ps pulses at the emitter resonance frequency. The emitted light is separated into two frequency bands using a dichroic mirror. The optical phonon sideband (PSB) at about 460~nm is collected into a single-mode fibre (``PSB arm'') conveying the photons to an avalanche photodiode (APD) for PLE characterisation, while the zero-phonon line (ZPL) is collected by another single-mode fibre (``ZPL arm''), which can be directed to various fibre-based characterisation setups. Laser rejection is implemented by setting the excitation arm polarisation to linear, with an axis at 45$^\circ$ relative to the emitter dipole direction. A second polariser, orthogonal to the excitation polarisation axis, is inserted in the collection arm, letting about half of the collected photons while blocking most of the reflected laser light. A quarter waveplate is placed in front of the objective to correct from residual ellipticity. Together with the polarising beamsplitter used to combine the excitation and collection modes, and the spatial mode filtering realised by the single-mode collection fibre, we obtain an extinction ratio of about $5\cdot 10^5$ in pulsed regime, and $8\cdot 10^5$ in~cw. { A detailed estimation of the collection, transmission and detection efficiency of our setup is available in the Supplementary Information section~S3.}

In the following, we focus on a single B centre { in the 220~nm crystal, emitting at 436.159~nm. We refer to this emitter as SPE$_1$. Its PL spectrum is provided in the Supplementary Information section~S1. All the experiments carried out with SPE$_1$ and described in the main text have been reproduced on another emitter, termed SPE$_2$, and the corresponding results are provided in Supplementary Information section S2.} The emitter properties are first inferred using PLE. For this characterisation, frequency filtering is sufficient to efficiently suppress the backreflected laser light. We excite the emitter using the cw tunable laser and collect photons in the PSB arm. On fig.~\ref{fig1}b, we show a low-power laser scan, revealing a linewidth of 0.95~GHz. This is about an order of magnitude larger than the Fourier limit of 93~MHz, due to the occurrence of discrete spectral jumps at micro- to millisecond timescales~\cite{Fournier23PRB, Gerard25}. In between spectral jumps, the coherence is maximal, \textit{i.e.} there is no appreciable pure dephasing. This is testified by the decay of Rabi oscillations observed in the cw second order correlation function $g^{(2)}(\tau)$ measured in the Hanbury Brown and Twiss (HBT) configuration, and shown fig.~\ref{fig1}c (red dots). The measurement is well fitted without requiring the introduction of pure dephasing processes (fig.~\ref{fig1}c, plain curve -- see Methods for the fitting procedure). We emphasize that, while the observation of Rabi oscillations allows us to characterize the electronic coherence in great detail, it does not provide any information about the photon coherence. Indeed, the emitter decay can involve coupling to phonons in such a way that the photons are highly incoherent. This is the case of many optically active defects in crystals, such as the nitrogen-vacancy centres in diamond~\cite{Ruf21}. In our case, the detection is performed using Stokes-shifted photons, which have a broad, red-detuned emission spectrum, and cannot be used for any quantum protocol based on photon interference. The coherence properties of the ZPL photons under resonant excitation will be addressed in the following sections.

\begin{figure*}[t]
  \centering
  \includegraphics[width=1.0\linewidth]{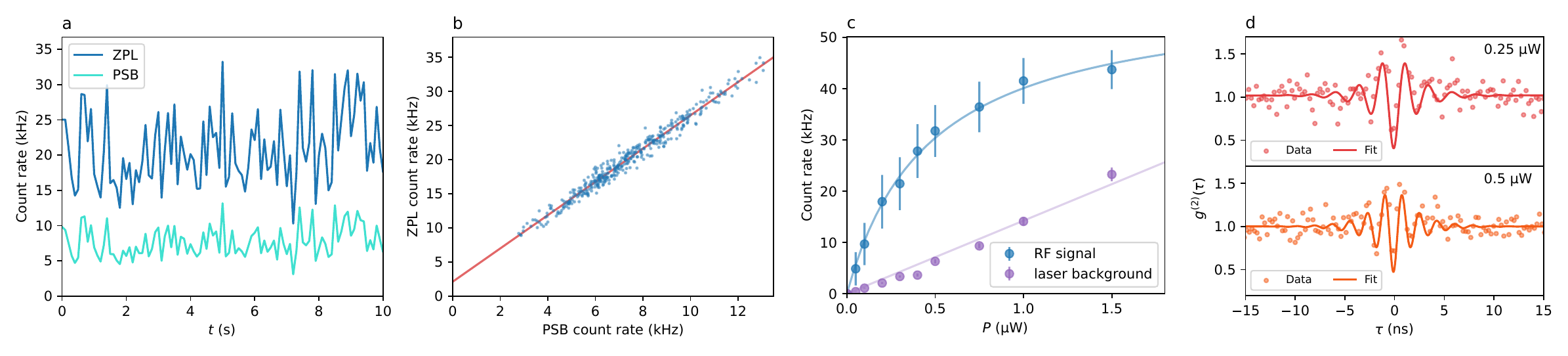}\\
  \caption{\textbf{Photon statistics of the ZPL signal.} (a) Intensity time traces in the ZPL (blue) and PSB (turquoise) detection arms. (b) Blue dots: ZPL count rate as a function of the PSB count rate in the same time bin. Red line: linear fit to the data. (c) Extracted emitter count rate (blue dots) and laser background (orange dots) as a function of the laser power. (d) Dots: $g^{(2)}(\tau)$ measured in the ZPL arm at four different powers, demonstrating Rabi oscillations. Plain lines: fit to the data.}  \label{fig2}
\end{figure*}

We then switch to the pulsed regime. Fig.~\ref{fig1}d shows the PSB count rate as a function of the pulsed laser power. {The count rate follows an oscillating dependence on the square root of the laser power, up to more than one period and a half. It can be fitted with a sine function with an exponential damping term of the form $\exp(-P/P_\mathrm{0})$, with $P_0 = 41$~$\mu W$. The damping is likely to originate from excitation-induced dephasing associated with acoustic phonons due to the large bandwidth of the laser~\cite{Ramsay10_1, Ramsay10_2}. The observation of these Rabi oscillations shows that the emitter can be deterministically prepared in the excited state using a $\pi$ pulse of average power 5~$\mu$W, with a fidelity of 95~\%.} This power is consistent with the cw Rabi oscillations, given the pulse length of 15~ps and the repetition rate of 76~MHz. We then measure the $g^{(2)}(\tau)$ in the pulsed regime under $\pi$ excitation to infer the emitter purity (blue curve on fig.~\ref{fig1}e). {A multi-peak fit (dashed black curve on fig.~\ref{fig1}e) provides $g^{(2)}(0) = 0.03 \pm 0.02$, corresponding to a purity of $0.97 \pm 0.02$ for the PSB photons}. This shows that the resonant pulsed excitation regime yields near ideal single-photon purity.

We now focus on light collected into the ZPL collection arm. The excitation and collection arm polarisers are set at 90$^\circ$ from each other, and at 45$^\circ$ from the emitter dipole axis. We tune the cw tunable laser on resonance with the emitter and we record the count rate from both the ZPL and the PSB collection arms. Fig.~\ref{fig2}a shows time traces measured simultaneously in the ZPL (blue curve) and the PSB (turquoise curve) arms, when the laser power is set low enough for the spectral diffusion to yield macroscopic (long-time) intensity fluctuations. These fluctuations take their origin from the fast ($<$~ms) frequency shifts, which translate to intensity fluctuations at the same time scale. An analysis based on the Mandel parameter shows that intensity fluctuations then persist at macroscopic times~\cite{Delteil24, Gerard25}. The synchronized fluctuations that can be observed suggest that the ZPL signal contains sizeable contribution from the emitter. To infer this contribution quantitatively, on fig.~\ref{fig2}b we plot the ZPL count rate as a function of the PSB count rate. The squeezed shape of the scatter plot demonstrates the strong correlation between intensity fluctuations in the two arms. A linear regression (red curve) allows us to extract the contribution of the laser light, which corresponds to the intercept -- \textit{i.e.} the remaining signal when the PSB count rate vanishes. The RF contribution is then obtained by subtracting the laser contribution from the average ZPL count rate. By repeating this process at various laser powers, we obtain the data shown fig.~\ref{fig2}c, where we plot both contributions as a function of the laser power. We observe the saturation of the RF contribution, while the laser contribution increases linearly. At low power, the SPE contribution is { a factor 8 higher than} the laser contribution. The ratio of RF over laser intensity is however not fundamentally limited to this amount. To further improve the SNR, several strategies can be considered. Firstly, the presence of spectral diffusion is detrimental to this ratio. At low power, the count rate is reduced by a factor $ \sqrt{\pi \ln 2} \Gamma / \Delta \omega_\mathrm{hom} \approx 6$, where $\Gamma$ is the emitter decay rate and $\Delta \omega_\mathrm{hom}$ the inhomogeneous linewidth~\cite{Gerard25}. This decrease is due to the reduced probability that the emitter is resonant with the laser at any given time. Improving the material quality, performing additional pre- or post-treatments~\cite{Horder25} or implementing electrical contacts~\cite{Akbari22} could strongly mitigate spectral diffusion. Additionally, photonic structures, such as cavities, could help reducing the saturation power so to decrease the laser contribution accordingly. Purcell enhancement of the ZPL can also increase the Debye-Waller factor and the collection efficiency by both decreasing the lifetime and optimizing the radiation pattern.

We then measure the $g^{(2)}(\tau)$ of the ZPL signal in the HBT configuration. The dots on Fig.~\ref{fig2}d show the results for four different powers. At the lowest powers, the antibunching dip is clearly visible. In addition, Rabi oscillations can also be observed. At higher power, the visibility decreases due to the dominant laser contribution. The curves are fitted in a similar way as the PSB signal (see Methods). We note that extracting the laser contribution from these measurements is not straightforward due to the presence of intensity fluctuations originating from spectral diffusion. The (maximal) amount of antibunching for constant, perfectly antibunched signal with uncorrelated background is $A_\mathrm{max} = S^2/(S + B)^2$, where $S$ (resp. $B$) is the signal (resp. background) contribution. However, for a fluctuating signal -- as is the case here --, this expression modifies to $A_\mathrm{max} = (S^2 + \Delta S^2)/(S +B)^2$, where $\Delta S$ is the standard deviation of the signal intensity distribution. This requires to separately characterise the amount of antibunching and the signal fluctuations. On the other hand, the linear regression approach directly provides the signal over background ratio with high accuracy in the case of a fluctuating signal, with a much lower integration time.
 
\begin{figure*}[t]
  \centering
  \includegraphics[width=1.0\linewidth]{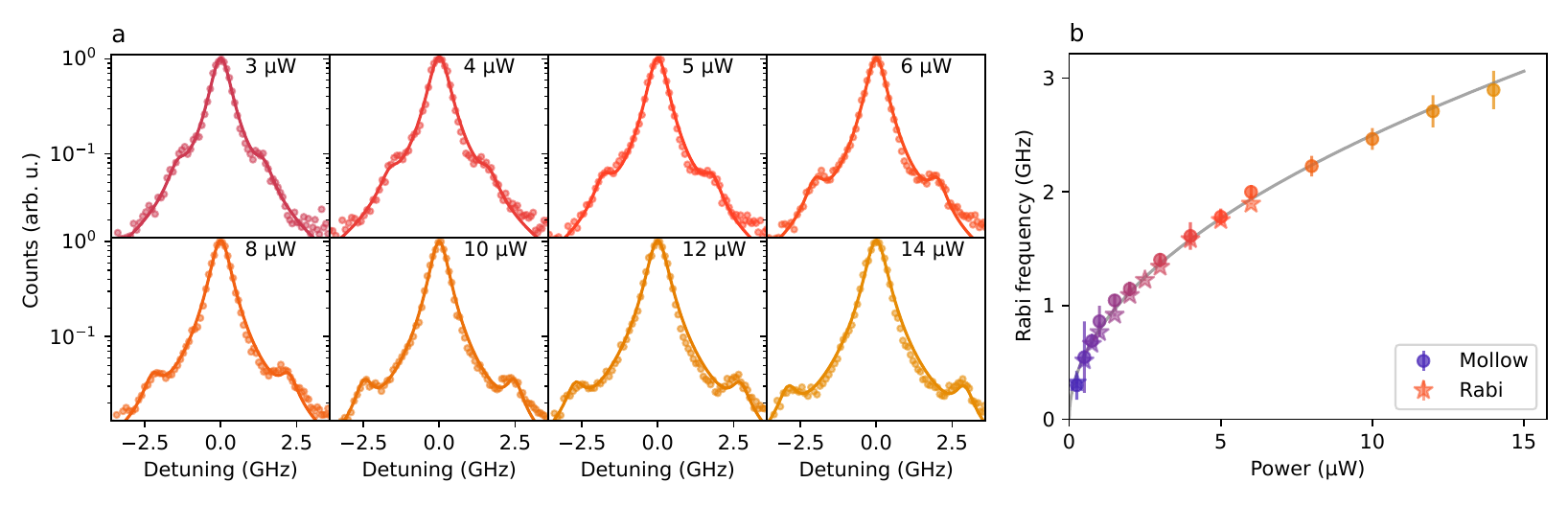}\\
  \caption{\textbf{Observation of the Mollow triplet.} (a) Dots: count rate as a function of the detuning of the scanning FP filter at height different powers. Plain line: fit to the data. (b) Circles: Rabi frequency extracted from the Mollow triplet splitting. Plain line: fit to the data. Stars: Rabi frequency extracted from the ZPL $g^{(2)}$ oscillations.}  \label{fig3}
\end{figure*}

The observation of SPE signal close to the laser frequency is not a conclusive proof of RF in the solid state, due to the presence of potential incoherent mechanisms associated with photon emission. To provide experimental indication that the ZPL signal is associated with atom-like RF, we perform high-resolution spectroscopy of the SPE emission. For this purpose, we couple the signal out of the collection fibre into a tuneable Fabry-Perot etalon of linewidth 0.75~GHz. Fig.~\ref{fig3}a shows spectra measured in the strong excitation regime, at laser powers varying from {3 to 14~$\mu$W}. While in this range, laser light dominates the collected signal, power-dependent symmetric sidebands can be observed on the spectra. These satellite peaks are associated with the Mollow triplet, a well-known signature of RF in the strongly driven regime~\cite{Mollow69}. In such conditions, the emitter is in strong coupling with the laser, resulting in split ground and excited states~\cite{Cohen98} The associated emission consists of four transitions, two of them being degenerate. Light emission in this regime therefore comprises a central peak at the laser energy and two satellites, spectrally shifted by $\pm \Omega_R$. We fit the data with three Lorentzian functions, with the constraint that the two outermost spectra are symmetric about zero detuning. The extracted splitting is plotted on fig.~\ref{fig3}b as a function of the laser power (dots on fig.~\ref{fig3}b). It can be well fitted by a square root function, in agreement with the expected theoretical dependence~\cite{Mollow69}. Additionally, we also report on the figure the values of the Rabi frequency extracted from the cw $g^{(2)}$ function (star symbols on fig.~\ref{fig3}b), confirming that the lateral peaks are the Mollow sidebands. This fingerprint of RF confirms that the ZPL emission takes place as a coherent light-matter interaction, where the emitted photons inherit the electronic coherence.

A natural expectation for maximally coherent emitters exhibiting coherent light emission -- and one of the main reasons for their quest -- is that they can in principle produce indistinguishable photons. The standard experimental benchmarking of photon indistinguishability is the demonstation of Hong-Ou-Mandel (HOM) interference~\cite{Couteau23}. This well-known effect corresponds to coalescence of photon pairs when made simultaneously incident on the two input ports of a beamsplitter, with all their internal degrees of freedom being identical. In the case of B centres, only the ZPL component is expected to exhibit the HOM effect, since the simultaneous emission of phonons during the emission process leaks which-path information, which prevents any interference. In hBN, HOM effect has already been observed in photoluminescence~\cite{Fournier23PRA}, but yielded partial indistinguishability of $0.55 \pm 0.11$, close to the classical limit of 0.5 due to the presence of dephasing. We attribute this limited visibility to a direct consequence of the non-resonant excitation regime, which requires high power, generates phonons, and increases the emitter spectral diffusion by a large amount. This reduced coherence under non-resonant excitation is also observed for other types of emitters, such as self-assembled quantum dots~\cite{Bennett05,Ates09}. For this reason, it is crucial to use resonant pulsed excitation to generate highly indistinguishable photons. To this end, we isolate ZPL photons while exciting resonantly using a frequency-doubled Ti:Sapphire laser. The pulse width of 100~GHz is much broader than the emitter inhomogeneous linewidth, such that the effects of spectral diffusion on the excitation efficiency are negligible. In addition to cross-polarisation suppression, we exploit the fact that the laser pulse is spectrally broader than the emitter transition linewidth to filter the light emission using a 10~GHz bandwidth Fabry-Perot etalon placed in the input port of the HOM interferometer. Additionally, we use the timescale mismatch between the laser pulse length and the emitter lifetime to perform { time gating of the remaining laser light based on post-selection of the detection events. The laser power is set to 5~$\mu$W, corresponding to a $\pi$ excitation.} Fig.~\ref{fig4}a shows a histogram of photon detection events in the pulsed, resonant regime. An intense, short peak can be observed at small delays. It originates from the remaining laser light, broadened by the instrument response function of 165~ps. At longer times, a weaker signal can be observed. It decays with a timescale of 1.70~ns, matching the B center decay time observed in the PSB port. The unsuppressed laser signal is also shown on Fig.~\ref{fig4}a (orange shading) for comparison. It does not exhibit any nanosecond-scale decay. The signal over noise ratio (SNR) is close to 20 at the start of the emitter decay, limited by dark counts and background signal, mainly from the PPLN crystal fluorescence. The latter limits the SNR that can be obtained based on this particular picosecond pulse generation technique, even with time gating.

\begin{figure}[t]
  \centering
  \includegraphics[width=0.85\linewidth]{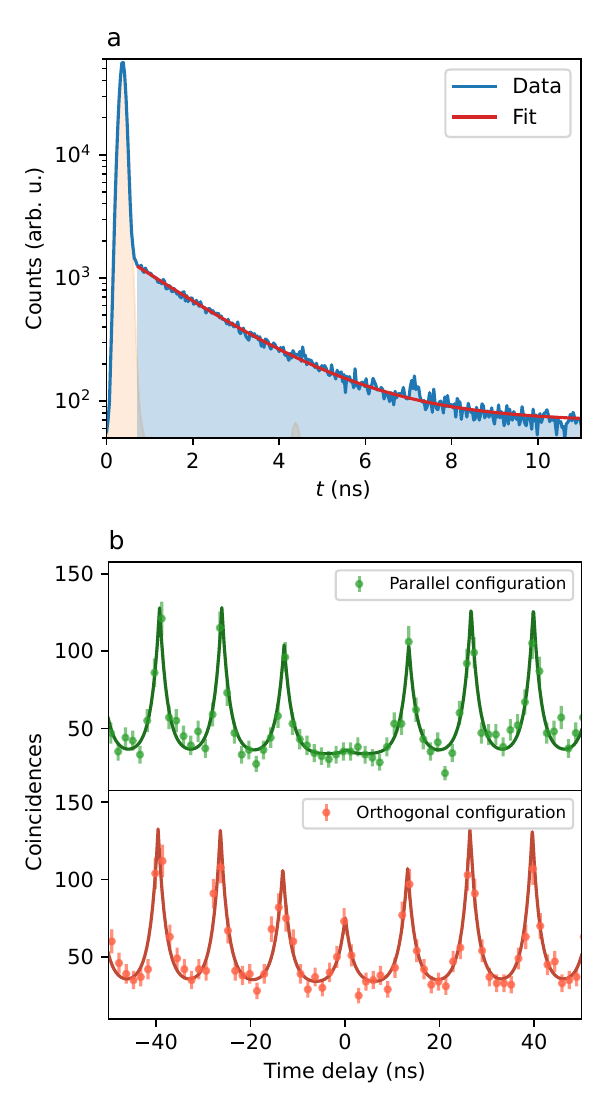}\\
  \caption{\textbf{Two-photon interference in the resonant pulsed regime.} (a) Blue line: histogram of the photon detections in the pulsed regime. Light from the laser is visible as the initial short peak, followed by a slower decay. The red line is a fit to the data, providing the relaxation time $T_1 = 1.70$~ns. Blue shading: events falling into the post-selection time window. Orange shading: unsuppressed laser, normalized to the same peak value. (b) Green dots: two-photon coincidences measured in the parallel configuration. Orange dots: coincidences measured in the orthogonal configuration. Plain lines: {multi-peak} fits to the data.}  \label{fig4}
\end{figure}

We then characterize the indistinguishability of the {post-selected} photons in the blue shaded region on fig.~\ref{fig4}a, \textit{i.e.} excluding the pulsed laser background. The HOM interferometer is depicted on fig.~\ref{fig1}a, and consists of a Mach-Zehnder interferometer of which one arm is delayed by the laser repetition period, such that consecutive photons can be simultaneously incident on the second beamsplitter. A liquid crystal retarder allows us to tune the polarisation of the two arms to be either parallel or orthogonal. Fig.~\ref{fig4}b displays the raw coincidence events measured at the output of the HOM interferometer for both the parallel polarisation case (green dots) and the orthogonal case (orange dots), each measured during 62~h. The dark curves are {multi-peak} fits to the data. { This fitting approach makes it possible to reliably extract the values of $g^{(2)}(0)$ even in the presence of sizeable uncorrelated background or partial peak overlap~\cite{Gazzano13, Somaschi16, Kirsanske17}} (see Methods). We extract $g^{(2)}_\parallel(0) = 0.03 \pm 0.09$ and  $g^{(2)}_\perp(0) = 0.44 \pm 0.08$ for the parallel, resp. orthogonal polarisation case. We deduce an estimation of $V_\mathrm{HOM} = 0.93 \pm 0.21$ { of the post-selected events}. We note that the upper bound of the HOM visibility that can be obtained in our experimental setup is given by the visibility of classical interference, which we measured to be $V_\mathrm{i} = 0.94 \pm 0.02$, and is likely the main limitation to the B centre HOM visibility. { The visibility of HOM interference between photons from the same source is expected to remain high as long as there is no spectral jump between photon emission events, \textit{i.e.} at delays shorter than a few milliseconds~\cite{Gerard25}. Interference between photons separated by longer times, as well as between photons from distinct sources, will require to mitigate spectral diffusion using improved material, electrical gating or Purcell enhancement. As mentioned earlier, all presented measurements have been repeated on another emitter (SPE$_2$) hosted in the 170~nm thick hBN crystal. The results are shown in the Supplementary Information section~S2. Although the SNR is lower for SPE$_2$ due to a less efficient cavity effect and suboptimal emitter position, resulting in a higher saturation power and a lower count rate, all the signatures of RF in cw and pulsed regime are well reproduced. The HOM visibility for SPE$_2$ is found to be $V_\mathrm{HOM} = 0.92 \pm 0.26$.} Altogether, these results demonstrate the potential of B centres for the generation of highly indistinguishable photons. 



In summary, we have implemented RF in both cw and pulsed regime on individual B centres, demonstrating that both laser excitation and photon emission are quantum coherent processes. This has allowed us to garner the best of these emitters in terms of photon indistinguishability. The low coincidence rate of filtered ZPL photons calls for an increased effort on integration in photonic structures, which could either be monolithic~\citep{Froch21, Nonahal23} or hybrid~\cite{Vogl19, Drawer23}. Beyond improving the SNR in cw regime, such photonic enhancement would lead to the generation of indistinguishable photons at high rates. Since B centres can be controllably positioned in photonic devices~\cite{Gerard23, Spencer23} and exhibit quasi identical spectra that can be electrically finely tuned~\cite{Zhigulin23}, our work opens the way to the implementation of quantum photonic processors based on identical emitters~\cite{Wan20}.


\section{Methods}
\subsection{Sample fabrication}

The hBN crystal is exfoliated from bulk material grown by the high-pressure, high-temperature method~\cite{Taniguchi07} onto a SiO$_2$(300~nm)/Si substrate. The B centres are created using 5~kV irradiation in a commercial scanning electron microscope. The crystal is then dry-transferred onto a Al$_2$O$_3$(1~nm)/Ag(80~nm)/SiO$_2$(300~nm)/Si substrate realised by radio frequency sputtering. The detailed fabrication method can be found in~\cite{Gerard24}.

\subsection{Optical measurements}

{ The two studied emitters are selected according to their count rate, saturation power, and dipole orientation. The latter falls into one of three main orientations separated by 60$^\circ$~\cite{Horder24}. The sample is oriented such that some emitters have a dipole orientation at about 45$^\circ$ relatively to the PBS axes, which is verified manually by realising emission extinction with a polariser inserted in the output arm.}

Resonant cw excitation is performed using an external cavity laser diode, frequency-locked to a wavemeter with resolution $\sim$~4~fm. Resonant pulsed excitation is performed using a femtosecond Ti:Sapphire laser, frequency-doubled using a periodically-poled lithium niobiate crystal allowing for second-harmonic generation with quasi phase matching. The laser extinction based on polarisation suppression is performed using a pair of double Glan-Taylor calcite polarisers in association with a polarising beamsplitter and a quarter waveplate. Photons are detected using avalanche photodiodes (Micro Photon Devices PDM-UV). { A time-correlated single-photon counter allows us to measure the photon count rate (for resonant laser scans and RF spectroscopy), the histogram of the delays between photon detections in start-stop mode (for cw and pulsed $g^{(2)}$ measurements), as well as to time-tag all detection events relatively to the laser pulse for time gating by post-selection, which is used in the HOM experiments. A detailed estimation of the collection, transmission and detection efficiencies of the whole setup is provided in the Supplementary Information section~S3.}


\subsection{Fitting procedure for the HBT and HOM $g^{(2)}$ functions}

\subsubsection{Fitting of cw $g^{(2)}$}

The cw HBT $g^{(2)}$ measurements shown Figs.~\ref{fig1}c and \ref{fig2}d are fitted using the following generalized function, valid for maximally coherent emitters subject to spectral diffusion~\cite{Fournier23PRB, Gerard25, Koch24}

\begin{equation}
\tilde{g}^{(2)}(\tau,\omega_L) = A \int d\omega S( \omega) C(\Omega_R,\omega_L, \omega)^2 g^{(2)}(\tau, \omega - \omega_L) + B
 \label{shorttimeg2}
\end{equation}
with $S(\omega)$ the inhomogeneous distribution (or probability density function) of the emitter center frequency $\omega$, $C(\Omega_R,\omega_L, \omega)$ the homogeneous response of the emitter, and $g^{(2)}(\tau, \omega - \omega_L)$ the homogeneous second order correlation function~\cite{Loudon}. In addition, we account for the presence of time-independent background by introducing $A$ the relative intensity of the signal and $B$ the background -- including the remaining laser background in RF~--, with $A + B = 1$.

\subsubsection{Fitting of pulsed $g^{(2)}$}

The HBT and HOM $g^{(2)}$ functions are fitted using {multi-peak} functions of the form
\begin{equation}
g^{(2)} = A \sum_{i \in \mathbb{Z}} c_{i} \exp(-|\tau - T_i|/T_1) + B
\end{equation}
where $T_i = iT_\mathrm{rep}$ with $T_\mathrm{rep}$ the laser period of 13.1~ns, $T_1 = 1.91$~ns the emitter lifetime, which is determined by fluorescence decay, and $c_{i}$ is the amplitude coefficient for the period $i$. $A$ and $B$ are fit parameters accounting for the signal and background intensities. For HBT measurements, $c_{i}$ is fixed to 1 except from the central period coefficient $c_0$, which is the fit parameter providing the value $g^{(2)}(0)$. For HOM, $c_{i}$ is fixed to 1 for $|i| \geq 2$, $c_{\pm 1} = 3/4$ and $c_{0}$ are fit parameters providing $g^{(2)}(0)$ for both parallel and orthogonal configurations. { In both cases, we consider periods with $i \leq 10$ for fitting the experimental data. The height of all lateral peaks is fixed to the same value, consistently with the absence of bunching in long-time $g^{(2)}$ under resonant excitation by a broad pulsed laser~\cite{Gerard25}. Note that $c_{0}$ identifies with the ratio of the center period area to the average of the areas of non-zero periods. Similarly, the ratio $g^{(2)}_\parallel(0)/g^{(2)}_\perp(0)$ used to infer the HOM visibility coincides with the ratio of the central peak areas for orthogonal and parallel polarisations.}

\section{Acknowledgments}
We thank B. Berini and A. Pierret for their contribution to sample fabrication. This work is supported by the French Agence Nationale de la Recherche (ANR) under reference ANR-21-CE47-0004. K.W. and T.T. acknowledge support from the JSPS KAKENHI (Grant Numbers 21H05233 and 23H02052), the CREST (JPMJCR24A5), JST and World Premier International Research Center Initiative (WPI), MEXT, Japan.

\section{Data availability}

The data generated in this study are available at ....

\pagebreak
~
\newpage

\onecolumngrid
\begin{center}
  \textbf{\large Resonance fluorescence and indistinguishable photons \\from a coherently driven B centre in hBN}\\[.2cm]
  Domitille G\'erard$^{1}$, St\'ephanie Buil$^{1}$, Kenji Watanabe$^{2}$, Takashi Taniguchi$^{3}$, Jean-Pierre Hermier$^{1}$, Aymeric Delteil$^{1}$\\[.1cm]
  {\itshape \small $^1$ Universit\'e Paris-Saclay, UVSQ, CNRS,  GEMaC, 78000, Versailles, France. \\
$^2$ Research Center for Electronic and Optical Materials, National Institute for Materials Science, 1-1 Namiki, Tsukuba 305-0044, Japan \\
$^3$ Research Center for Materials Nanoarchitectonics, National Institute for Materials Science,  1-1 Namiki, Tsukuba 305-0044, Japan \\ {\color{white}--------------------} aymeric.delteil@usvq.fr{\color{white}--------------------} \\}

\end{center}

\setcounter{equation}{0}
\setcounter{figure}{0}
\setcounter{table}{0}
\setcounter{page}{1}
\renewcommand{\theequation}{S\arabic{equation}}
\renewcommand{\thefigure}{S\arabic{figure}}

\section{S1. Photoluminescence characterization}

\begin{figure}[h]
  \centering
  \includegraphics[width=0.5\linewidth]{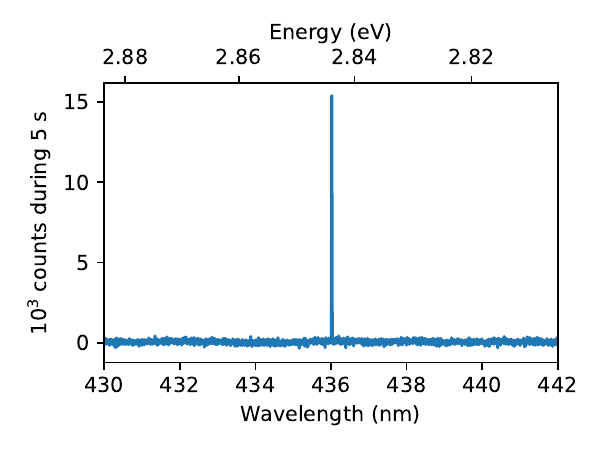}\\
  \caption{PL spectrum of SPE$_1$.  \label{fig0}}
\end{figure}

Fig.~\ref{fig0} shows a photoluminescence spectrum of the emitter studied in the main text (SPE$_1$), measured with a fibre-coupled echelle spectrometer. A single peak is observed at about 436~nm. Its linewidth is limited by the spectrometer resolution (50~$\mu$eV).

\section{S2. Charaterization of an additional B centre}

We repeat the whole set of measurement described in the main text for another B centre (termed SPE$_2$) emitting at 436.018~nm. The sample structure is identical to that studied in the main text (SPE$_1$), except for the crystal thickness (170~nm) and the vertical position of the emitter, which is uncontrolled. This results in a lower enhancement of the collection efficiency and a higher saturation power~\cite{Gerard24}. Fig.~\ref{fig1} shows the PLE characterisation in both cw and pulsed regimes. In the cw regime, the resonant laser scan on Fig.~\ref{fig1}a provides the time-integrated linewidth of 0.75~GHz. The $g^{(2)}(\tau)$ shown on Fig.~\ref{fig1}b allows us to verify that the emitter is maximally coherent at timescales shorter than the spectral diffusion time. In the pulsed regime, the count rate as a function of the laser power exhibits the onset of Rabi oscillations, with a $\pi$ pulse power of 60~$\mu$W, sizeably higher than for SPE$_1$. Finally, the $g^{(2)}$ in pulsed regime (Fig.~\ref{fig1}d) yields $g^{(2)}(0) = 0.03 \pm 0.03$.

\begin{figure}[h!]
  \centering
  \includegraphics[width=0.9\linewidth]{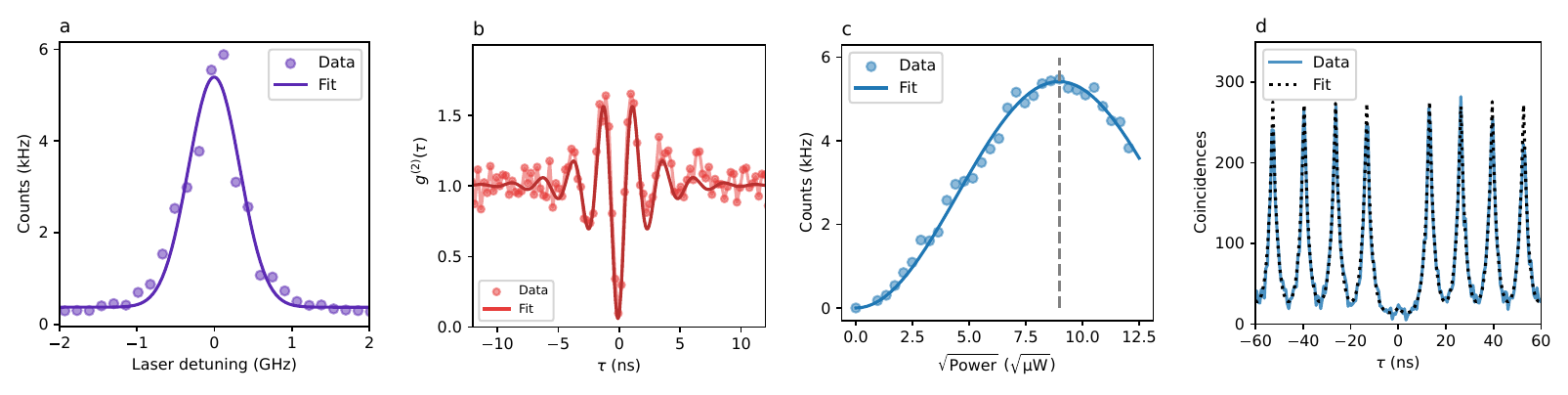}\\
  \caption{\textbf{PLE characterisation of the emitter.} (a) Dots: low-power (1.5~$\mu$W) laser scan to infer the emitter inhomogeneous linewidth. Plain line: Gaussian fit to the data. (b) Dots: medium-power (5~$\mu$W) $g^{(2)}(\tau)$ revealing Rabi oscillations. Plain line: fit to the data. (c) Dots: count rate as a function of the square root of the laser power in pulsed regime. Plain line: sine fit to the data. (d) Blue curve: $g^{(2)}(\tau)$ in pulsed regime. Dotted line: multi-peak fit to the data.  \label{fig1}}
\end{figure}

The time traces of the zero-phonon line under low-power cw excitation exhibit correlated fluctuations similarly to SPE$_1$ (Fig.~\ref{fig2}a). A similar statistical analysis (Fig.~\ref{fig2}b) provides the SNR. The result is shown on Fig.~\ref{fig2}c. In the case of SPE$_2$, the count is comparable to the laser background at low power, and exhibit saturations at powers of a few microwatts. Measurements of the $g^{(2)}$ function at various powers reveal Rabi oscillations.

\begin{figure}[h!]
  \centering
  \includegraphics[width=1.0\linewidth]{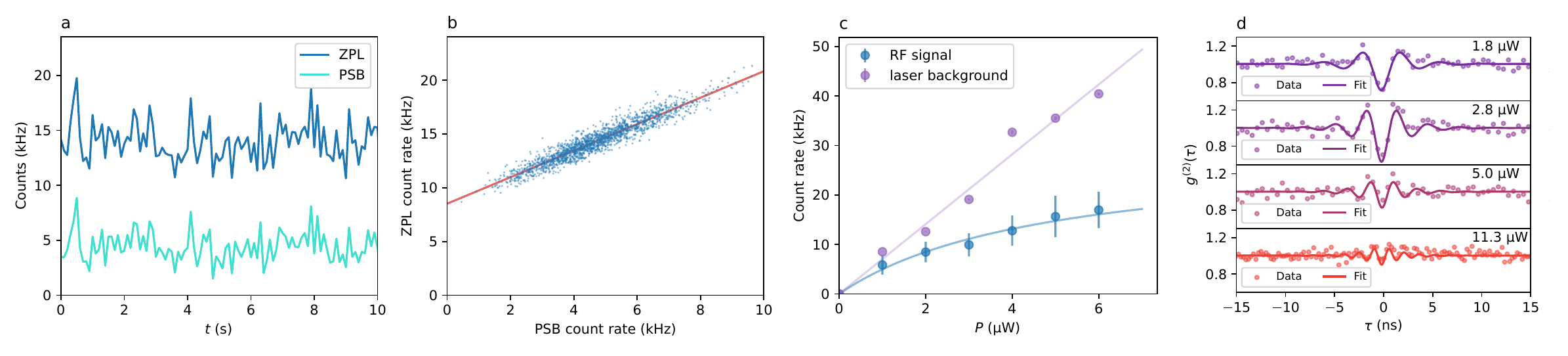}\\
  \caption{\textbf{Photon statistics of the ZPL signal.} (a) Intensity time traces in the ZPL (blue) and PSB (turquoise) detection arms. (b) Blue dots: ZPL count rate as a function of the PSB count rate in the same time bin. Red line: linear fit to the data. (c) Extracted emitter count rate (blue dots) and laser background (orange dots) as a function of the laser power. (d) Dots: $g^{(2)}(\tau)$ measured in the ZPL arm at four different powers, demonstrating Rabi oscillations. Plain lines: fit to the data.}  \label{fig2}
\end{figure}

Spectroscopy of the ZPL emission in the high power regime allows us to observe Mollow sidebands with a power-dependent splitting, as can be seen on Fig.~\ref{fig3}a. The measurement and fit processes are identical to those performed on SPE$_1$, and allow us to confirm the square root dependence of the splitting as a function of the laser power (circles on Fig.~\ref{fig3}b), which is in good agreement with the Rabi frequency extracted from $g^{(2)}(\tau)$ measurements (stars on Fig.~\ref{fig3}b).
 
\begin{figure}[h!]
  \centering
  \includegraphics[width=1.0\linewidth]{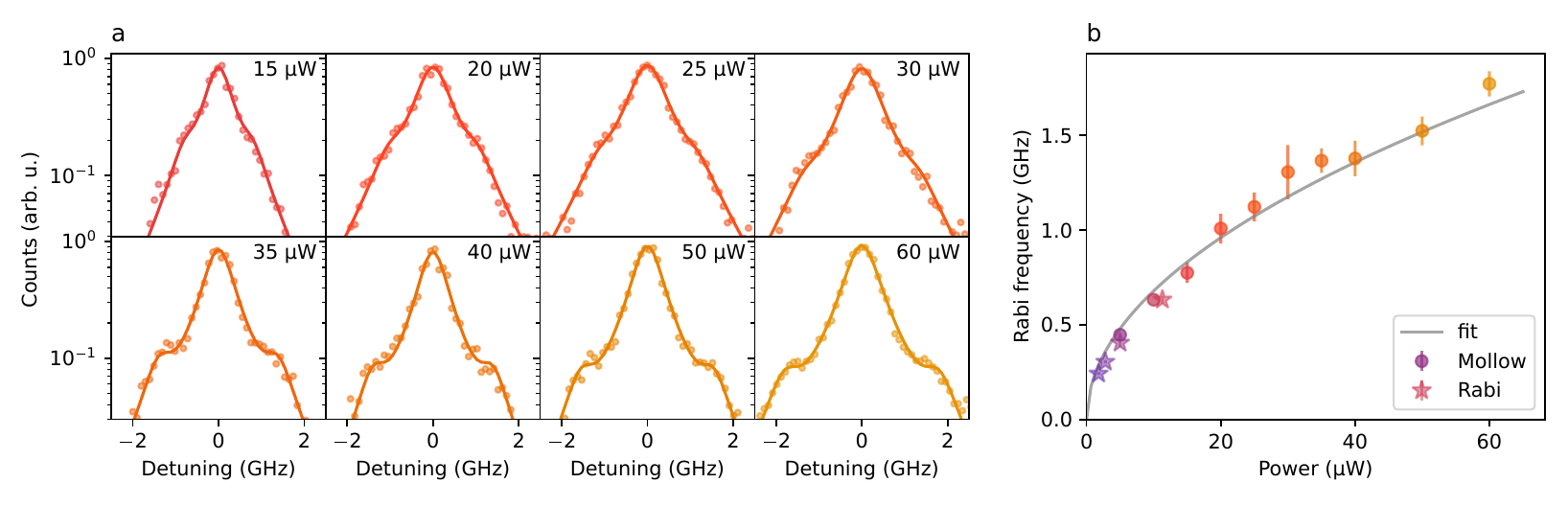}\\
  \caption{\textbf{Observation of the Mollow triplet.} (a) Dots: count rate as a function of the detuning of the scanning FP filter at height different powers. Plain line: fit to the data. (b) Circles: Rabi frequency extracted from the Mollow triplet splitting. Plain line: fit to the data. Stars: Rabi frequency extracted from the ZPL $g^{(2)}$ oscillations.}  \label{fig3}
\end{figure}

Fig.~\ref{fig4} shows the HOM interference measurement in pulsed regime. The laser power is set at 70~\% of a $\pi$ pulse (corresponding to an excited state population of 0.92), as a trade-off between optimising the count rate and minimising the laser background. The RF decay is shown on Fig.~\ref{fig4}a can be used to identify the laser background (orange shading) and the SPE emission (blue shading), which we post-select in the HOM measurement. The green (resp. orange) dots Fig.~\ref{fig4}b shows the coincidence histogram in parallel (resp. orthogonal) polarisation configuration, each measured during 180~h. A multi-peak fit provides the values of $g^{(2)}_\parallel(0) = 0.04 \pm 0.12$ and  $g^{(2)}_\perp(0) = 0.45 \pm 0.11$ for the parallel, resp. orthogonal polarisation case. We deduce an estimation of $V_\mathrm{HOM} = 0.92 \pm 0.26$, in good agreement with the values obtained for SPE$_1$.

\begin{figure}[h!]
  \centering
  \includegraphics[width=0.7\linewidth]{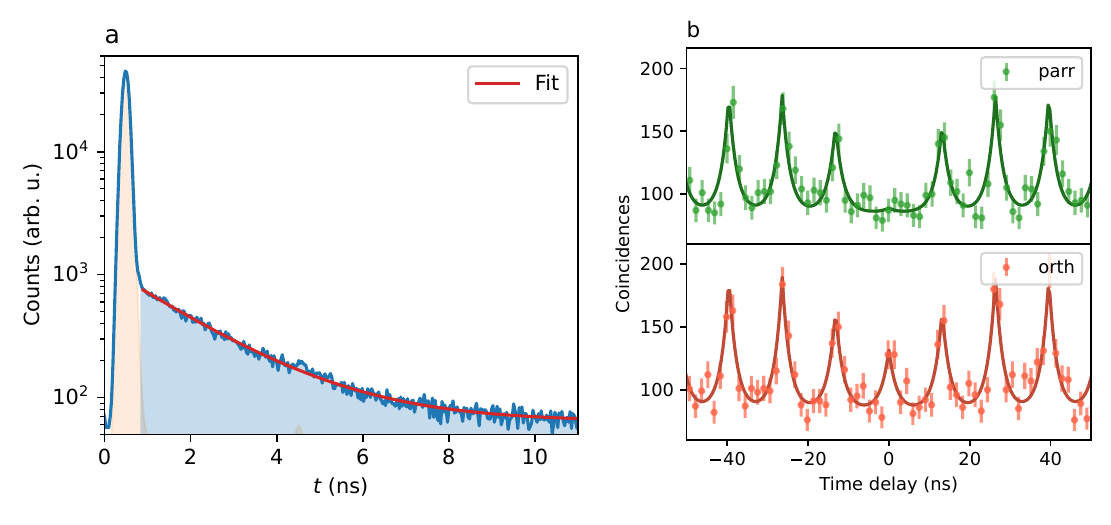}\\
  \caption{\textbf{Two-photon interference in the resonant pulsed regime.} (a) Blue line: histogram of the photon detections in the pulsed regime. Light from the laser is visible as the initial short peak, followed by a slower decay. The red line is a fit to the data, providing the relaxation time $T_1 = 1.91$~ns. Blue shading: events falling into the post-selection time window. Orange shading: unsuppressed laser, normalized to the same peak value. (b) Green dots: two-photon coincidences measured in the parallel configuration. Orange dots: coincidences measured in the orthogonal configuration. Plain lines: multi-peak fits to the data.}  \label{fig4}
\end{figure}


\section{S3. Count rate and setup collection and detection efficiency}

In the Hong-Ou-Mandel experiment, we measure about 250~coincidences per 13.1~ns period over 62~h integration time, corresponding to a count rate of 600~Hz on the two APDs under $\pi$ excitation at $f_\mathrm{rep} = 76$~MHz -- \textit{i.e.} a photon detection probability of about $8 \times 10^{-6}$ per pulse. This low count rate is explained by the cumulated losses in the different stages of the setup. In the following, we provide an estimation of the collection, transmission and detection efficiencies of the various parts of our setup.\\

\textbf{Transmission of the cross-polarisation setup $T_\mathrm{CP}$.} The output count rate after the filtering setup for separating the photon emission from the laser light is about 8~\% of the original count rate. This is due to the following reasons:

-- The dichroic mirror removes about 20~\% of the total counts, corresponding to the LO phonon replica, which is collected in the other arm.

-- The cross-polarisation suppression removes 50~\% of the light emission

-- The transmission of the double Glan-Taylor is about 50~\% in the relevant wavelength range.

-- The coupling efficiency to the output fibre (also including fibre losses) is about 50~\%.\\

\textbf{Transmission of the HOM interferometer $T_\mathrm{interf}$.} The interferometer has a total transmission of about 4~\%. This includes:

-- Grating peak transmission: 65~\%

-- Fabry-Perot filter peak transmission: 80~\%

-- Fraction of the coherent ZPL photons within the near-resonant light emission: 40~\%.

-- Cumulated transmission of the other optical elements (waveplates, polarizer): 50~\%.

-- Fibre coupling, insertion losses and fibre connectors: 40~\%
\\

\textbf{APD efficiency $\eta_\mathrm{APD}$.} The detection efficiency of the APDs is about 40~\%. \\

\textbf{Collection efficiency $\eta_\mathrm{coll}$.} The collection efficiency is defined as the proportion of emitted photons that we collect with our microscope objective and convey to our free-space setup. It includes the fraction of the light which is emitted towards the top of the sample, the efficiency of the collection by the microscope objective, and the transmission of the objective and of the cryostat windows. We can estimate it from two of the experiments presented in the main text.

-- The saturation rate $R_\mathrm{sat}$ of the ZPL photons of 60~kHz (Fig.~2c) is measured through the cross-polarisation setup and detected by an APD, thus $R_\mathrm{sat} = 0.5 \Gamma_1 \eta_\mathrm{coll}T_\mathrm{CP}\eta_\mathrm{APD}$, where the initial factor 0.5 stands for the excited state population above saturation. From this, we estimate $\eta_\mathrm{coll} \approx 0.75$~\%.

-- The measurement of Rabi oscillations in the pulsed regime (Fig.~1d) leads to a count rate of about 8~kHz in the PSB arm under $\pi$ excitation at 76~MHz repetition rate. The upper bound of the PSB arm throughput $T_\mathrm{PSB}$ is about 0.5. In addition, the optical PSB emission is only about 15~\% of the total spectrum, and the polarised beamsplitter transmission is 0.5 for 45$^\circ$ polarized light. Accounting for the APD detection efficiency, we obtain $\eta_\mathrm{coll} \approx 0.7$~\%, consistently with the estimation from the saturation count rate.\\

The count rate $R$ at the output of the HOM interferometer can be estimated as $R = f_\mathrm{rep} \eta_\mathrm{coll} T_\mathrm{CP} T_\mathrm{interf} \eta_\mathrm{APD} \approx 680$~Hz, which is consistent with the measured count rate. This also confirms our estimation of the collection efficiency, which could be improved using photonic structures such as 0D cavities. The planar structure used in this work, while improving the count rate by a factor $\sim 2$ in average compared with a bare exfoliated crystal~\cite{Gerard24}, is insufficient to yield a high collection efficiency, as would a structure with lateral light confinement, such as DBR-based micropillars, bullseye cavities, etc. due to the predominance of emission to in-plane propagating modes, which we do not collect. We also note that the internal quantum efficiency of the B centres was estimated to be close to unity using the same samples~\cite{Gerard24}, and measurements of long-time $g^{(2)}$ under pulsed excitation have excluded the presence of efficient shelving states~\cite{Gerard25}, which would have further degraded the count rate.

\end{document}